\documentclass[twoside]{LCWS11}
\usepackage[latin1]{inputenc}
\usepackage[dvips]{graphicx,epsfig,color}
\usepackage{wrapfig,rotating}
\usepackage{amssymb,amsmath,array}
\usepackage{cite}
\newcommand{\be}{\begin{equation}}
\newcommand{\ee}{\end{equation}}
\newcommand{\bea}{\begin{eqnarray}}
\newcommand{\eea}{\end{eqnarray}}

\def\ssc{\scriptscriptstyle}
\def\lsim{\mathrel{\raise.3ex\hbox{$<$\kern-.75em\lower1ex\hbox{$\sim$}}} }
\def\gsim{\mathrel{\raise.3ex\hbox{$>$\kern-.75em\lower1ex\hbox{$\sim$}}} }

\begin{document}
%%%%%%%%%%%%%%%%%%%%%%%%%% net cover %%%%%%%%%%%%%%%%%%%%%%%%%%
\thispagestyle{empty}

\begin{flushright}
NCU-HEP-k049  \\
Jan 2012
\end{flushright}

\vspace*{.5in}

\begin{center}
{\bf Dynamical Symmetry Breaking in  Supersymmetric Extensions of
Nambu--Jona-Lasinio Model}\\
\vspace*{.5in}
{
{\bf
Gaber Faisel $^{1,2}$ Dong-Won Jung $^3$ Otto C. W. Kong $^1$}
% Optional short acknowledgment: remove next line if non-needed
\vspace{.3cm}\\
% Addresses and institutions (remove "1- " in case of a single institution)
1- Department of Physics and \\
Center for Mathematics and Theoretical Physics,\\ National Central
University, Chung-Li, Taiwan 32054.
%% Remove the next three lines in case of a single institution
\vspace{.1cm}\\
2- Egyptian Center for Theoretical Physics,\\
Modern University for Information and Technology, Cairo, Egypt.
\vspace{.1cm}\\
3- Department of Physics, National Tsing Hua University and \\
Physics Division, National Center for Theoretical Sciences,
Hsinchu, Taiwan, 300\\
}

\vspace*{1.in}
\end{center}
{\bf Abstract :}\ \
%\abstracts
{In this paper we discuss Nambu--Jona-Lasinio model as a classical
model for  dynamical mass generation and symmetry breaking. In
addition we discuss the possible supersymmetric extensions of this
model  resulting from interaction terms with four chiral
superfields that may be regarded as a supersymmetric
generalization of the four-fermion interactions of the
Nambu--Jona-Lasinio model. A four-superfield interaction terms can
be constructed as either dimension 6 or dimension 5 operators.
Through analyzing solutions to the gap equations, we discuss the
dynamical generation of superfield Dirac mass, including a
supersymmetry breaking part. A dynamical symmetry breaking
generally goes along with the dynamical mass generation, for which
a bi-superfield condensate is responsible.
 }
%\end{center}
%\twocolumn[\maketitle\abstract{
%}]

\vfill
\noindent --------------- \\
$^\star$ Talk presented by Gaber Faisel at LCWS 2011 (September
26-30), Granada, Spain.\\
 --- submission for the proceedings.

\clearpage
\addtocounter{page}{-1}
%%%%%%%%%%%%%%%%%%%%%%%%%% net cover ends %%%%%%%%%%%%%%%%%%%%%%%%%%

\title{
%%%%   Paper title goes here  %%%%%%%%%%%%%%
 Dynamical Symmetry Breaking in  Supersymmetric Extensions of
Nambu--Jona-Lasinio Model } %%
%***********************************************************************
% AUTHORS INFORMATION AREA
%***********************************************************************
\author{Gaber Faisel $^{1,2}$ Dong-Won Jung $^3$ Otto C. W. Kong $^1$
%\thanks{}
% DO NOT MODIFY THE FOLLOWING '\vspace' ARGUMENT
\vspace{.3cm}\\
% Addresses and institutions (remove "1- " in case of a single institution)
1- Department of Physics and \\
Center for Mathematics and Theoretical Physics,\\ National Central
University, Chung-Li, Taiwan 32054.
%% Remove the next three lines in case of a single institution
\vspace{.1cm}\\
2- Egyptian Center for Theoretical Physics,\\
Modern University for Information and Technology, Cairo, Egypt.
\vspace{.1cm}\\
3- Department of Physics, National Tsing Hua University and \\
Physics Division, National Center for Theoretical Sciences,
Hsinchu, Taiwan, 300\\
}
%%***********************************************************************
% END OF AUTHORS INFORMATION AREA
%***********************************************************************

\maketitle

\begin{abstract}

 In this paper we discuss Nambu--Jona-Lasinio model
as a classical model for  dynamical mass generation and symmetry
breaking. In addition we discuss the possible supersymmetric
extensions of this model  resulting from interaction terms with
four chiral superfields that may be regarded as a supersymmetric
generalization of the four-fermion interactions of the
Nambu--Jona-Lasinio model. A four-superfield interaction terms can
be constructed as either dimension 6 or dimension 5 operators.
Through analyzing solutions to the gap equations, we discuss the
dynamical generation of superfield Dirac mass, including a
supersymmetry breaking part. A dynamical symmetry breaking
generally goes along with the dynamical mass generation, for which
a bi-superfield condensate is responsible.
\end{abstract}

\section{Introduction}

  Nambu adopted the idea of Cooper pairing
\cite{Nobel} to construct a classic model of dynamical mass
generation and symmetry breaking. This is the Nambu--Jona-Lasinio
(NJL) model \cite{Nambu:1961tp}, with a strong attractive
four-fermi interaction.   After the Standard Model was generally
established, the exact mechanism of electroweak symmetry breaking
became a problem of paramount importance in the phenomenological
domain. It is still open till today. It was pointed out by Nambu
\cite{Nambu} that for a sufficiently heavy top quark, an NJL model
of top condensate can give rise to electroweak symmetry breaking.
The top quark, however, turns out to be not heavy enough
\cite{tSM,Chung:2005yz}.

\begin{figure}
\centerline{\includegraphics[width=0.45\columnwidth]{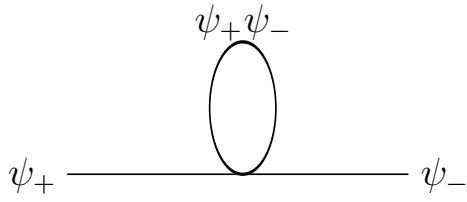}}
\caption{ Diagram for proper self-energy
$\Sigma_{+-}(p)$.\label{smprop}}
\end{figure}

The Lagrangian for the NJL model can be written as  \bea {\mathcal
L}&=&i\partial_m\psi_+\sigma^m\bar{\psi}_++i\partial_m\psi_-\sigma^m\bar{\psi}_-
+ g^2\psi_+\psi_-\bar{\psi}_+\bar{\psi}_- \label{LSM}\eea here
$\psi_+,\psi_-$ are two-component Weyl spinors and the coupling g
has mass dimension -1 which shows that the model is to be taken as
an effective theory and has to be provided with a cut-off
$\Lambda$. Clearly from eq.(\ref{LSM}),
 the Lagrangian ${\mathcal L}$ is invariant under the chiral $U(1)$
transformations:

\bea U(1)_V&:&  \psi_{\pm}\to e^{\pm i\alpha} \psi_{\pm}
\nonumber\\U(1)_A&:& \psi_{\pm}\to e^{ i\beta} \psi_{\pm} \eea We
can rewrite ${\mathcal L}$ as

\be {\mathcal L}={\mathcal L}_0+{\mathcal L}_I\ee

where \be {\mathcal
L}_0=i\partial_m\psi_+\sigma^m\bar{\psi}_++i\partial_m\psi_-\sigma^m\bar{\psi}_-
-m \psi_-\psi_+
 \ee
and the interaction Lagrangian ${\mathcal L}_I$ is given by \be
{\mathcal L}_I=g^2\psi_+\psi_-\bar{\psi}_+\bar{\psi}_- +m
\psi_-\psi_+\ee The mass $m$ is self-consistently defined by \be
\Gamma^2(p)_{\gamma p=-m}=0 \ee where $\Gamma^2(p)$ is the proper
two- point function. This yields the gap equation:

\be \label{SMg} m = \left. \Sigma^{({\mbox \tiny loop})}_{+-}(p)
 \right|_{\mbox{\tiny on-shell}} \;,
\ee where $\Sigma^{({\mbox \tiny loop})}_{+-}$ denotes
contributions from the proper self-energy  diagram  shown in
Fig.(\ref{smprop}). Upon evaluation of the diagram, the gap
equation
 reads \be m= m g^2\frac{\Lambda^2}{8\pi^2} \left[ 1
- \frac{|m|^2}{\Lambda^2} \, \ln{\frac{\Lambda^2 }{|m|^2}} +
O(1/\Lambda^4) \right] \;.\ee which has nontrivial solution i.e.
$m\neq 0$ for  coupling constant satisfying the inequality \be g^2
> \frac{8\pi^2}{N_C\Lambda^2} \label{SMgs} \;, \ee  Clearly,  the
nonperturbative gap equation, eq.(\ref{SMg}), with eq.(\ref{SMgs})
show that the  strong attractive four-fermi interaction  induces a
bi-fermion vacuum condensate of the operator $\psi_+\psi_-$ which
serves as the source for the fermion Dirac mass. Moreover, the
condensate naturally breaks  the chiral symmetry that the
bi-fermion carries, which was Nambu's first concern
\cite{Nobel,N60}.

\section{Supersymmetric extensions of NJL model\label{gapsec}}

 A supersymmetric extension of the NJL model via  dimension six four-superfield
interaction (SNJL) was introduced in 1982 \cite{BL}.  The  gap
equation analysis showed no nontrivial mass solution. The model
can be recovered leading to dynamical mass generation upon
introducing soft supersymmetric breaking mass terms \cite{BE}.
However, phenomenological viability of the model has been severely
unfavorable cornered with the relatively small top mass value
determined and constraint on the $\tan\!\beta$ parameter
\cite{034}. A  natural alternative to SNJL model as elaborated in
Ref.\cite{Kong:2010zza} was presented in Ref.\cite{034}, together
with an explicit model version that can give rise to electroweak
symmetry breaking. The new model has a dimension five
four-superfield interaction in the superpotential and hence it is
holomorphic and named as (HSNJL) model. Recently a fully detailed
study for SNJL and HSNJL models based on introduction of a new
perspective on the superfield gap equation using the supergraph
technique has been presented in ref.\cite{Faisel:2011te}.   The
explicit illustration of dynamical symmetry breaking from HSNJL
showed rich and novel features, which would be easily missed
without the superfield approach developed there.

\begin{figure}
\centerline{\includegraphics[width=0.45\columnwidth]{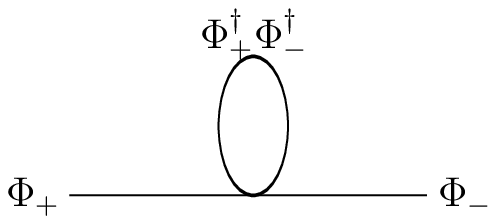}
\includegraphics[width=0.45\columnwidth]{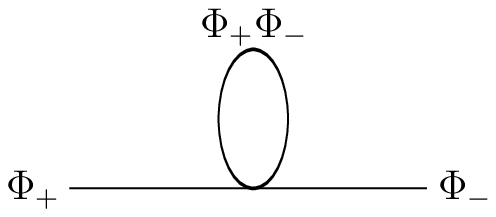}}
\caption{ Supergraphs for proper self-energy $\Sigma_{+-}(p,
\theta^2)$. Left (right) supergraph corresponds to
 dimension six (five) four-superfield interaction.\label{susgraph}}
\end{figure}

The key point in the analysis of ref.\cite{Faisel:2011te} is
extending the  gap equation for the Dirac mass $m$,
eq.(\ref{SMg}), to  \be \label{gap} - {\mathcal M} = \left.
\Sigma^{({\mbox \tiny loop})}_{+-}(p,\theta^2)
 \right|_{\mbox{\tiny on-shell}} \;,\ee
 where ${\mathcal M}$ is given  in  ${\cal L}_I$ which is written in terms of
the chiral superfields that contain $\psi_+$ and $\psi_-$ as one
of their components. ${\mathcal M}$  contains  the usual
(supersymmetric) Dirac mass $m$  and its supersymmetry breaking
counterpart $\eta$ . The former corresponds to Dirac mass for the
fermion pair $\psi_\pm$ and $|m|^2$ contributions to both $A_\pm$
mass-squared, while the supersymmetry breaking part $\eta$ gives
(so-called left-right) mass mixing between the $A_\pm$ pair. In
eq.(\ref{gap}) $\Sigma^{({\mbox \tiny loop})}_{+-}$ denotes
contributions from the proper self-energy diagram involving the
interaction. The interactions of interest that are expected to
lead to nontrivial $\Sigma^{({\mbox \tiny loop})}_{+-}$ can be
given by the dimension six four-superfield interaction  \be
\label{d6} g^2 \int\!\! d^4 \theta \, \Phi_+^\dagger\Phi_-^\dagger
\Phi_+\Phi_- \, (1- \tilde{m}_{\!\ssc C}^2 \theta^2 {\bar
\theta}^2 ) \ee coming with a supersymmetry breaking part which
gives the SNJL model, here extended to include the supersymmetry
breaking $\tilde{m}_{\ssc C}^2$ part. The HSNJL model  alternative
has rather a dimension five four-superfield interaction which is
given by \be \label{d5} -\frac{G}{2} \int\!\! d^4 \theta \,
\Phi_+\Phi_-\Phi_+\Phi_- \, (1+ B \theta^2)\,
\delta^2\!(\bar{\theta}) \;. \ee It is really a superpotential
term, as indicated by the $\delta^2\!(\bar{\theta})$, hence
holomorphic.

\subsection{Dimension six interaction}
The $g^2$ vertex gives at one-loop level the proper self-energy
diagram shown in Fig.(\ref{susgraph}) left. The gap equation,
eq.(\ref{gap}) with $\Sigma^{({\mbox \tiny fig2})}$, reads
\cite{Faisel:2011te} \bea
 m &=& 2 m  g^2  I_1(|m|^2, \tilde{m}^2, |\eta|, \Lambda^2) \;,
\nonumber \\
\eta &=& - \eta \, g^2 \tilde{m}_{\!\ssc C}^2 \, {I}_2(|m|^2,
\tilde{m}^2, |\eta|, \Lambda^2) \;. \label{gap6} \eea

where \bea I_1(|m|^2, \tilde{m}^2, |\eta|, \Lambda^2) &=&
\frac{1}{16\pi^2} \left[ \frac{1}{2}(|m|^2+\tilde{m}^2)
\ln{\frac{(|m|^2+\tilde{m}^2+\Lambda^2)^2-|\eta|^2}{(|m|^2+\tilde{m}^2)^2-|\eta|^2}}
-|m|^2 \ln{\frac{(|m|^2+\Lambda^2)}{|m|^2}} \right. \nonumber \\
 && + \left. |\eta|
\left(\tanh^{-1}\frac{|m|^2+\tilde{m}^2+\Lambda^2}{|\eta|}
-\tanh^{-1}\frac{|m|^2+\tilde{m}^2}{|\eta|} \right) \right] \;,
\nonumber \\
I_2(|m|^2, \tilde{m}^2, |\eta|, \Lambda^2) &=&
 \frac{1}{16\pi^2} \left[ \frac{1}{2}
\ln{\frac{(|m|^2+\tilde{m}^2+\Lambda^2)^2-|\eta|^2}{(|m|^2+\tilde{m}^2)^2-|\eta|^2}}
\right. \nonumber \\
&& \!\!\!\! + \left. \frac{|m|^2+\tilde{m}^2}{|\eta|}
\left(\tanh^{-1}\frac{|m|^2+\tilde{m}^2+\Lambda^2}{|\eta|}
-\tanh^{-1}\frac{|m|^2+\tilde{m}^2}{|\eta|} \right) \right] .
\label{I12} \eea

Note that the case with both $\tilde{m}_{\!\ssc C}^2$ and
$\tilde{m}^2$ being zero corresponds to the SNJL model with an
exactly supersymmetric Lagrangian \cite{BL}. In that case, a
supergraph analysis has been performed going to two-loop
evaluation of $\Sigma^{({\mbox \tiny loop})}_{+-}(p,\theta)$
\cite{BL}. No nontrivial solution for $m$ exists. It should be
noted that when our result of eq.(\ref{gap6}) is applied to the
case, nontrivial solution for $\eta$ will imply spontaneous
supersymmetry breaking exist.

On the other hand, taking the limit  $\tilde{m}\rightarrow \infty$
where the scalar particles of $\Phi_{\pm}$ become heavy and
decoupled,  $m$ becomes the simple Dirac fermion/quark mass which
then satisfies eq.(\ref{SMg}) after including a factor $N_c$ to
account for the number of the  colors of the quarks.  A nontrivial
solution for $m$ exists for the coupling constant satisfying the
inequality \cite{BE} \be g^2
> \frac{8\pi^2}{\tilde{m}^2 \ln \left(
1+\frac{\Lambda^2}{\tilde{m}^2} \right)}   \;, \ee generating a
mass for the Dirac fermion pair.

Considering  the scenario $m=0$ but $\eta \ne 0$ solution for
Eq.(\ref{gap6}). Naively, one enforces zero $m$ in the the $I_2$
integral of the equation for $\eta$. Nontrivial solution for the
latter exists under the condition \bea \frac{1}{16\pi^2}
\left[\ln{\left(1+\frac{\Lambda^2}{ \tilde{m}^2}\right)}
-\frac{\Lambda^2}{\Lambda^2+\tilde{m}^2}\right] \leq
\frac{1}{-g^2\tilde{m}_{\!\ssc C}^2} <
\frac{1}{16\pi^2}\ln{\left(1+\frac{\Lambda^2}{2\tilde{m}^2}\right)}
\;, \eea  The last part of the inequality comes from an analysis
similar to that of the condition for nonotrivial $m$ under
$\eta=0$. The magnitude of the responsible coupling,
${g^2\tilde{m}_{\!\ssc C}^2}$ here, has to be big enough. The
other part of the inequality is actually from $|\eta| \leq
(m^2+\tilde{m}^2)$ beyond which there will be a tachyonic scalar
mass eigenvalue. Note that one always needs a negative
${\tilde{m}_{\!\ssc C}^2}$ for nontrivial $\eta$ solution.

\subsection{Dimension five interaction}
Turning now to  the dimension five interaction case in which the G
vertex gives at one-loop level a diagram only slightly different
from the previous case, as shown in  Fig.(\ref{susgraph}) right.
The gap equation, eq.(\ref{gap}) with $\Sigma^{({\mbox \tiny
fig2})}$, reads \cite{Faisel:2011te} \bea m &=& \frac{\bar{\eta}
G}{2}\; I_2(|m|^2,\tilde{m}^2,|\eta|, \Lambda^2)\;,
\nonumber \\
\eta &=&  \bar{m} G \;  I_1(|m|^2,\tilde{m}^2,|\eta|, \Lambda^2) -
\frac{\bar{\eta} G B}{2} \; I_2(|m|^2,\tilde{m}^2,|\eta|,
\Lambda^2) \;. \eea %%

The first thing to note in the gap equation result is the
important fact that the equations for $m$ and $\eta$ are
completely coupled. If one naively drop $\eta$ from consideration,
for instance, one will not see any nontrivial expression and
completely miss the possible dynamical mass generation. The two
parameters will either both have nontrivial solutions or both
vanishing.

Considering only the case of real values for $m$ and $\eta$ under
the assumption of a real and small $B$ value, we find that
nontrivial solution exists for large enough G (taken as real and
positive here by convention) satisfying \be G > \sqrt{G_0^2 + b^2}
+ b \sim G_0  + b\;, \ee where \bea G_0^2 =
\frac{512\pi^2}{\tilde{m}^2\ln{\left(1+\frac{\Lambda^2}{\tilde{m}^2}\right)}
              \left[\ln{\left(1+\frac{\Lambda^2}{\tilde{m}^2}\right)}
 -\frac{\Lambda^2}{\tilde{m}^2+\Lambda^2}\right]} \;
\eea gives the critical $G^2$ for $B=0$, and \be b = B \;
\frac{8\pi^2}{\tilde{m}^2\ln{\left(1+\frac{\Lambda^2}{\tilde{m}^2}\right)}}
\;. \ee Details can be found in ref.\cite{Faisel:2011te}. Solution
condition for more general cases is to be further investigated.

%\begin{wrapfigure}{r}{0.5\columnwidth}
%\begin{figure}
%\centerline{\includegraphics[width=0.45\columnwidth]{samplefigure.eps}}
%\caption{UGR Logo}\label{Fig:MV}
%\end{wrapfigure}
%\end{figure}

\section{Conclusion}

In this talk we have discussed NJL model with its possible
supersymmetric extensions  that can be constructed from either
dimension 6 or dimension 5 four-superfield interactions. The two
kinds of four-superfield interactions may be considered
alternative supersymmetrization of the four-fermion interaction in
the NJL model of dynamical mass generation and symmetry breaking.
They could each be used as a mechanism for dynamical electroweak
symmetry breaking. The two kinds of models (SNJL and HSNJL models)
have otherwise very different theoretical mass generation
features, with phenomenological implications. In addition, we
presented  the superfield gap equations  for both dimension six
and  dimension five four-superfield interactions and discussed
some interesting cases for nontrivial solution.

 We have shown  also that  dimension five four-superfield interaction
can induce  the dynamical mass generation for the prototype HSNJL
model. The model has actually no four-fermion interaction and has
a bi-scalar condensate, instead of bi-fermion condensate, as the
source of Dirac fermion mass. It has otherwise theoretical
features that look like a direct supersymmetric version of the NJL
model.  It is expected to provide an alternative paradigm for
dynamical mass generation and symmetry breaking, at least for
superfield theories. The explicit symmetry breaking picture of the
simplest HSNJL model maybe considered as $Z_4 \rightarrow Z_2$. A
version of the HSNJL with the basics superfields being (gauge)
multiplets gives a simple application to the breaking of a
continuous symmetry.  The model can be also extended with more
than two basic superfield multiplets that can achieve a rich
spectrum of dynamical symmetry breaking. A case example, which was
also the original target for the idea of the HSNJL model is the
one for electroweak symmetry breaking which we refer to
ref.\cite{Faisel:2011te} for more details.

\section{Acknowledgments}

G.F. and O.K. are partially supported by research grant NSC 99-
2112-M-008-003-MY3, and G.F. further supported by grant NSC
100-2811-M-008-036 from the National Science Council of Taiwan.
Gaber Faisel  would like to thank Taiwan LHC Focus Group (NCTS)
for their financial support to participate in the  conference and
present this talk.

% ****************************************************************************
% BIBLIOGRAPHY AREA
% ****************************************************************************

% ****************************************************************************
% BIBLIOGRAPHY AREA
% ****************************************************************************

\begin{footnotesize}
% IF YOU DO NOT USE BIBTEX, USE THE FOLLOWING SAMPLE SCHEME FOR THE REFERENCES
% ----------------------------------------------------------------------------

% ----------------------------------------------------------------------------
\end{footnotesize}

% ****************************************************************************
% END OF BIBLIOGRAPHY AREA
% ****************************************************************************

\end{document}